\documentclass[a4paper]{jpconf}
\usepackage{graphicx}
\usepackage{amsmath}
\begin{document}
\title{Initial Conditions with Flow from a McLerran Venugopalan model with
Transverse Dynamics}

\author{Guangyao Chen and Rainer J.~Fries}

\address{Cyclotron Institute and Department of Physics and Astronomy, Texas A\&M University, College Station, TX 77843, USA}

\ead{glorychern@neo.tamu.edu}

\begin{abstract}
Using a recursive solution of the Yang-Mills equation, we calculate analytic
expressions for the gluon fields created in ultra-relativistic heavy ion
collisions at small times $\tau$. We have worked out explicit solutions for
the fields and the energy momentum tensor up to 4th order in an expansion in $\tau$.
We generalize the McLerran-Venugopalan model to allow for a systematic
treatment of averaged charge densities $\mu^2$ that vary as a function of
transverse coordinates. This allows us to calculate radial, elliptic and
directed flow of gluon fields. Our results can serve as initial conditions for hydrodynamic simulations of nuclear collisions that include initial flow.
\end{abstract}

\section{Introduction}

Nuclear collisions at high energies at the Relativistic Heavy Ion Collider
(RHIC) and the Large Hadron Collider (LHC) can create and probe novel forms of QCD
matter. One example is a deconfined phase of QCD called Quark Gluon Plasma
(QGP). A thermalized QGP is created in nuclear collisions at RHIC and LHC after a
time of about 1 fm/$c$. Relativistic hydrodynamics is the tool of choice to
describe the evolution of QGP in this
stage \cite{Adams:2005dq,PHENIX}. Large uncertainties in the hydrodynamic
evolution come from the incomplete understanding of the
initial conditions, i.e.\ the evolution of the nuclear collision prior to 1
fm/$c$.
This is where another new phase of QCD, Color Glass Condensates (CGC) plays an important part. The natural limit of a nucleus at very high energy is characterized
by a saturated gluon density in which the gluon field can be described by
a quasi-classical field \cite{McLerran:1993ni,McLerran:1993ka}. The earliest
phase of the collision of two nuclei can thus be described as the collision of
two sheets of color glass \cite{KoMLeWei:95}. The classical field resulting
from the collision then decays and equilibrates to a plasma of quarks and gluons.

To be more precise, in the color glass formalism partons are
distinguished by momentum fraction. Sources have large Bjorken-$x$ while the
small-$x$ gluons are treated through the classical field $F^{\mu\nu}$. They are related by
the Yang-Mills equation $[D_\mu, F^{\mu \nu}]=J^\nu$. Before the collision,
the two incoming nuclei on the positive and negative light cone are dominated
by transverse chromo-electric and chromo-magnetic fields which
are perpendicular to each other \cite{McLerran:1993ni,McLerran:1993ka}.
After the collision, the fields in the forward light cone can be parameterized as
$A^{\pm}= \pm x^{\pm} A$ and a transverse field $A^i$, $i=1,2$. $A$ and $A^i$ depend on the proper time
$\tau=\sqrt{t^2-z^2}$ and transverse coordinates but not on the space-time
rapidity $\eta$.  For small $\tau$, a series expansion $A^{\mu}=\sum_n \tau^n
A^{\mu}_{(n)}$ in powers of $\tau$ can be employed. A few lowest orders in
time should give an adequate and analytic description of the gluon fields immediately
after the collision \cite{FKL:06}.

The collision initially ($\tau \rightarrow 0$ ) creates longitudinal chromo-electric and magnetic fields
\begin{equation}
E_0=ig[A_1^i,A_2^i],  \qquad  B_0=-ig \epsilon^{ij} [A_1^i,A_2^j]   \nonumber
\end{equation}
between the nuclei, where $A_1^i$ and $A_2^i$ are the fields of nucleus 1 or 2
before the collision in light cone gauge. The next order in $\tau$ gives us
linearly growing transverse electric and magnetic fields
\begin{equation}
  E^i_{(1)} = -\frac{1}{2} ( \sinh\eta [D^i, E_0] + \cosh\eta \,
  \epsilon^{ij}[D^j,B_0] ),   \quad
  B^i_{(1)} = \frac{1}{2} ( \cosh\eta \, \epsilon^{ij} [D^j, E_0]
  - \sinh\eta [D^i,B_0] ).  \nonumber
\end{equation}
We have calculated analytic expressions for the fields up to fourth order in $\tau$
explicitly and will discuss those results elsewhere \cite{Chen}. The behavior
we find is qualitatively consistent with numerical studies of classical field
QCD \cite{Lappi:06}.

The energy momentum tensor of the gluon field after the collision can be
easily calculated. Up to order $\mathcal{O}(\tau^2)$ --- we skip higher orders
here for brevity --- it can be written in the form
\begin{equation}
  T^{m n}_{\mathrm f} =
  \begin{pmatrix}
    A+C  & B_1 & B_2  & 0 \\
    B_1  & A+D & E & B_1'/\tau \\
    B_2  & E & A-D & B_2'/\tau \\
    0    & B_1'/\tau & B_2'/\tau  & (-A +C)/\tau^2
      \end{pmatrix}
      \label{tmn}
\end{equation}
in the $\tau,x,y,\eta$ coordinate system.
Here $A=(E_0^2+B_0^2)/2$ is the initial energy density at $\tau=0$, and $B_i$ and
$B_i'$ are the coefficients of the Poynting vector $T^{0i}=B_i\cosh \eta
+B_i'\sinh \eta$ and linear in $\tau$ ($i=1,2$ are transverse indices). $C$, $D$, $E\propto \tau^2$
are coefficients quadratic in $\tau$. All these coefficients have been
calculated analytically \cite{Chen}.
One can check that this energy momentum tensor is boost invariant owing to
the original boost-invariant setup of the colliding nuclei.

\section{Flow in the McLerran Venugopalan Model}

In the McLerran-Venugopalan (MV) model \cite{McLerran:1993ni,McLerran:1993ka},
the simplest implementation of color glass, an observable $O$ calculated from
sources in the two nuclei has to be averaged over all possible configurations of
color source distributions $\rho$. Each collision samples arbitrary $\rho$ due
to the short time scale of the collision compared to internal time scales of
the nuclear wave function.
The distribution of charge is usually assumed to be Gaussian around zero with
an average value $\mu^2$  of the variation. We use the definition
\begin{align}
\langle \rho_a (\vec x_\perp)\rho_b (\vec y_\perp) \rangle =
\frac{g^2 \delta_{ab}}{N_c^2-1} \mu^2(\vec x_\perp) \delta^2(\vec x_\perp-\vec y_\perp)
\, .
\end{align}
We have suppressed the dependence on the longitudinal coordinate
for brevity. Note that we have generalized the original expression from the MV
model such that $\mu^2$ can be a function of transverse coordinate $\vec
x_\perp$, instead of being an average charge density that is assumed to be homogeneous.
Of course the latter is unrealistic and the generalization needs to be made if
long-range transverse dynamics is to be described.
We have shown explicitly that if $\mu^2$ is varying slowly
such that $|\mu^2(\vec x_\perp)| \gg m^{-1} |\nabla^i \mu^2(\vec x_\perp)|
\gg m^{-2} |\nabla^i \nabla^j \mu^2(\vec x_\perp)| $, color glass dynamics and
long-distance dynamics which are not described by CGC can be safely separated.
Here $m$ is an infrared regulator. We find that the generalized two gluon
correlation function is \cite{Chen}
\begin{equation}
\langle A_a^i(\vec x_\perp)A_b^j(\vec x_\perp)\rangle = \delta_{ab} \frac{g^2 \mu^2(\vec x_\perp)}{8 \pi (N_c^2-1)} \big[ \delta^{ij} \ln \frac{Q^2}{(1.47m)^2} + \frac{ \nabla^k \nabla^l \mu^2(\vec x_\perp)}{m^2 \mu^2(\vec x_\perp)}(\frac{1}{6} \delta^{kl}\delta^{ij} - \frac{7}{12} \delta^{ik}\delta^{jl}) \big],
\end{equation}
where the first term is the same as in the original MV model \cite{JMKMW:96}. $Q$
here is the saturation scale.

\begin{figure}[tb]
\begin{minipage}{11pc}
\includegraphics[width=11pc]{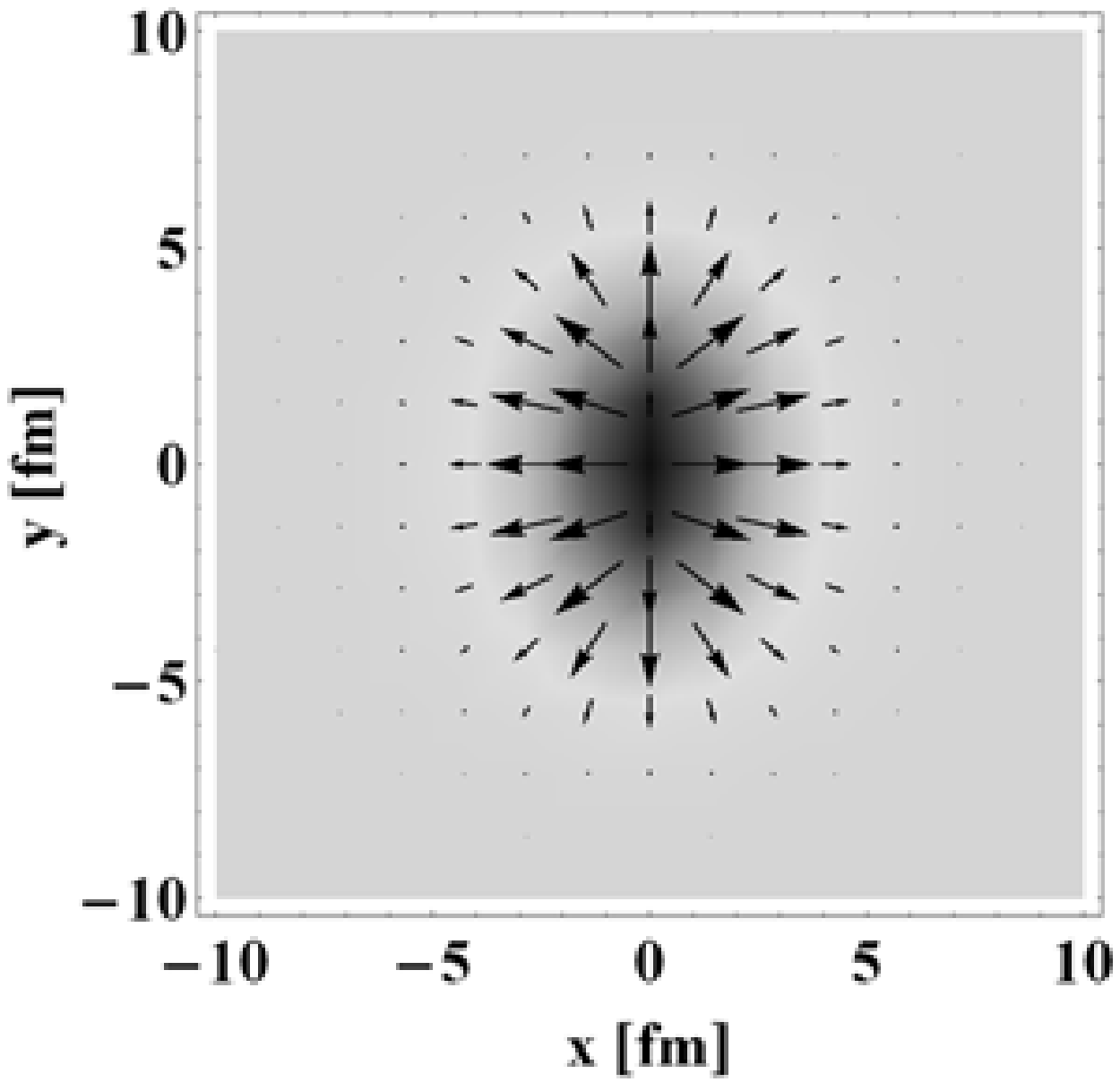}
\end{minipage}\hspace{2pc}%
\begin{minipage}{11pc}
\includegraphics[width=11pc]{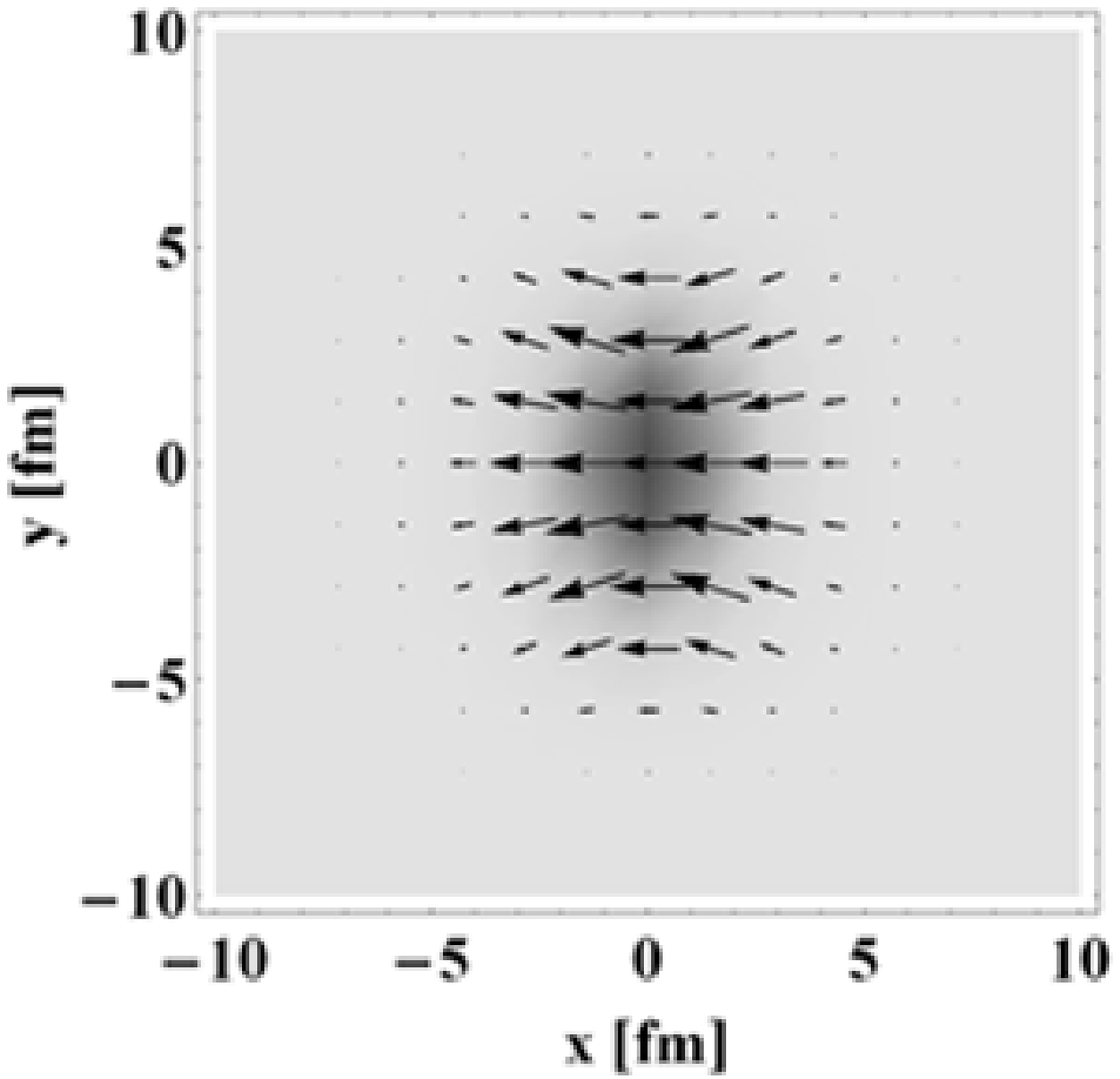}
\end{minipage}\hspace{2pc}%
\begin{minipage}{11pc}
\includegraphics[width=11pc]{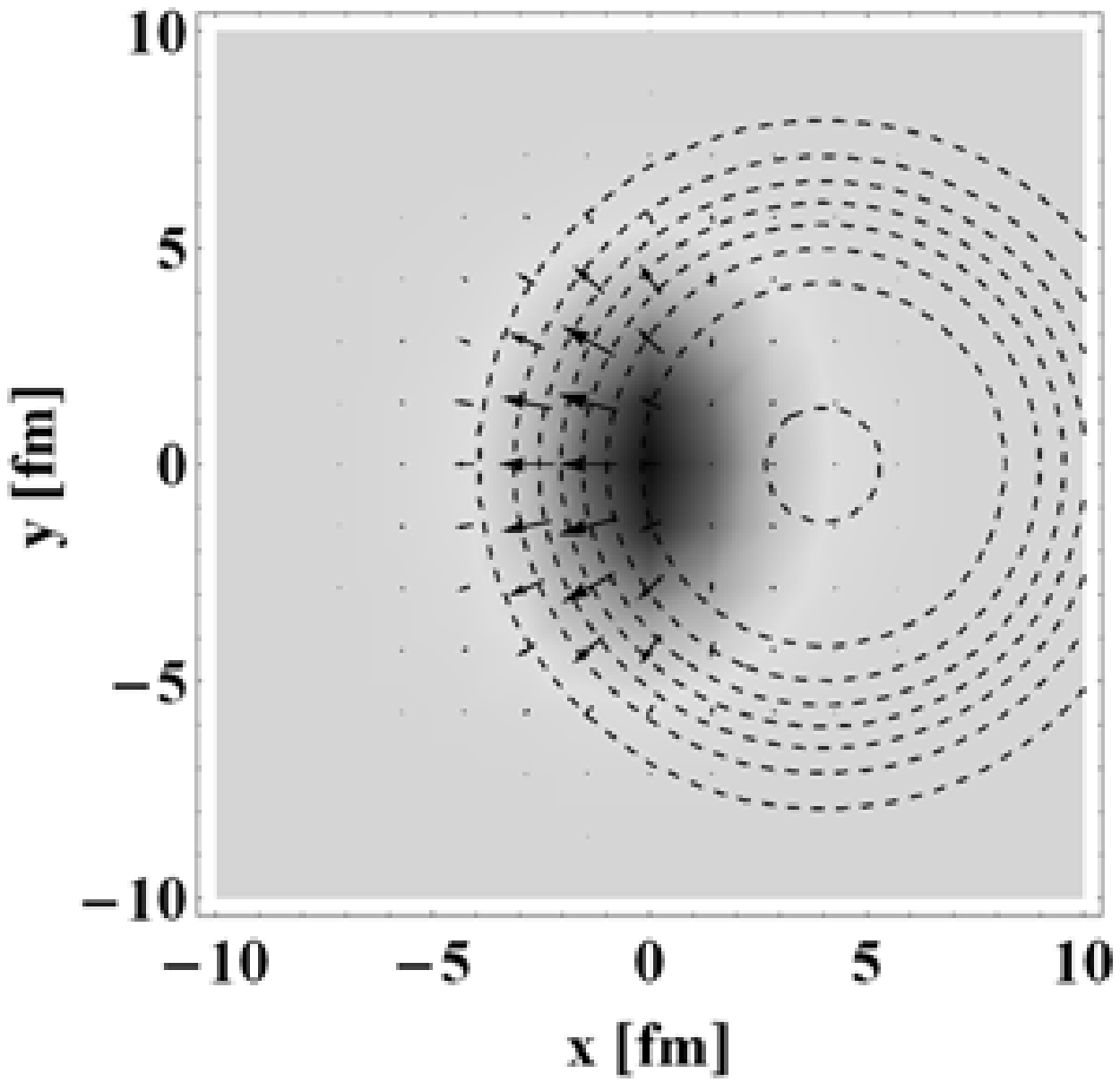}
\end{minipage}
\caption{\label{fig:1} Left: The hydro-like flow component $\vec B$ (arrows) and the
 underlying energy density. Middle: The rapidity-odd flow component $\vec B'$. Right: The Poynting vector $T^{0i}$ at $\eta=1$ together with
  the energy density. The contour lines indicate the density profile of one of
the nuclei.}
\end{figure}

Let us now focus on the transverse Poynting vector $T^{0i}$ in the energy momentum
tensor which describes the energy flow of the gluon field.
The term even in $\eta$ after averaging reads
\begin{align}
B_i=- \tau \frac{g^6 N_c (N_c^2-1) }{32 \pi} \nabla^i (\mu^2_1 \mu^2_2) \ln^2 \frac{Q^2}{(1.47m)^2}
\, .
\end{align}
This expression gives radial and elliptic flow of energy $\propto
-\nabla^i \epsilon$. This is similar to what a hydrodynamic flow field would look like for the same energy density $\epsilon$ if the hydrodynamic pressure is a monotonous function of $\epsilon$.
The term odd in $\eta$ is
\begin{align}
B_i'=-\tau \frac{g^6 N_c (N_c^2-1) }{32 \pi}  \ln^2 \frac{Q^2}{(1.47m)^2}
[\mu^2_2 \nabla^i \mu^2_1 - \mu^2_2 \nabla^i \mu^2_1] \, .
\end{align}
Interestingly this term can describe a directed flow of energy. Note that this
term does not violate boost-invariance of the energy momentum tensor. On the
other hand boost-invariant hydrodynamics would never be able to create such a flow field. Rather one can trace this term back to the QCD analogon of Gauss' Law.
This makes this rapidity-odd flow field a curious phenomenon.
The flow vector fields $\vec B$ and $\vec B'$ from two gold nuclei with Woods-Saxon
profiles $\mu^2$  colliding with impact parameter $b=8$ \,fm are shown in
Fig.\ \ref{fig:1}.  The net Poynting vector $T^{0i}$
exhibits a clear dipole asymmetry in the transverse plane when $\eta \ne0$.
Flow in color glass at midrapidity has recently also been studied numerically in \cite{Schenke:2012wb}.

\section{Hydrodynamic Initial Conditions}

Since the classical gluon field only describes the very earliest stage of a
nuclear collision we need to translate the energy momentum tensor into initial
conditions for hydrodynamics. The dynamics of equilibration is outside the
purview of either hydrodynamics nor classical field theory and would require a
more complete description of quantum non-abelian dynamics. One can however
derive a matching hydrodynamic energy momentum tensor simply using conserved
currents both on the classical field side and the hydrodynamic side. For ideal hydrodynamics this was worked out in \cite{Fries:2005yc, Fries:2007iy} by just imposing energy and
momentum conservation. The flow of energy generally translates into a flow of
particles on the hydrodynamic side. The direction of the hydrodynamic
transverse flow velocity is parallel to the Poynting vector and its size is a
monotonous function of the magnitude of the Poynting vector.  A similar
correspondence can be worked out in the case of matching to viscous
hydrodynamics \cite{Chen}.

\section{Directed Flow Phenomenology}

Without a further hydrodynamic evolution we can nevertheless check wether our
somewhat surprising direct flow result is qualitatively consistent with
experiments. The first thing to check is the direction of the flow. It turns
out that our directed flow vector $\vec B'$ points away from the spectators
for a given side $\eta>0$ or $\eta<0$ of a collision system with finite impact
parameter. This is consistent with experiments. The rapidity dependence of
particle directed flow $v_1$ is also roughly consistent with a $\sinh y$ shape
as a function of particle rapidity $y$ \cite{Abelev:2008jga,Adams:2005ca}.

Note that $\vec B'$ vanishes for $b=0$ if the two nuclei are
equal. Interestingly, if an asymmetric system like Cu+Au collides with $b=0$
the flow vector $\vec B'$ becomes radial, but points outward in the direction
of the smaller nucleus and inward in the direction of the larger nucleus. An
asymmetric system with $b\ne 0$ would exhibit a striking blend of both
effects. Looking for such a strong dependence of flow on rapidity in
central collisions of asymmetric systems could be a signature for the eminence
of strong gluon fields early in nuclear collisions.

\section{Conclusion}

To summarize, if the McLerran Venugopalan model is generalized to allow for
variations of the average charge density across the nucleus flow phenomena in
early stages of nuclear collisions can be studied. Here we use analytic
results for early gluon fields and their energy momentum tensor that can be obtained
in an expansion of the Yang-Mills equations around the time of collision.
We find radial and elliptic flow patterns similar to hydrodynamic behavior. In addition
we also obtain a rapidity-odd term in the flow of gluon energy which resembles
directed flow. We have argued that this flow will translate into directed flow
of particles in the further course of the collision. The results are in
qualitative agreement with results from the STAR experiment.

\ack{This work is supported by NSF CAREER grant PHY-0847538, JET Collaboration
 and DOE grant DE-FG02-10ER41682. We thank Joseph Kapusta and Yang Li who
contributed to this project.}

\end{document}